# Wasm-iCARE: a portable and privacy-preserving web module to build, validate, and apply absolute risk models

Jeya Balaji Balasubramanian[1], Parichoy Pal Choudhury[1,2], Srijon Mukhopadhyay[1], Thomas Ahearn[1], Nilanjan Chatterjee[1,3], Montserrat García-Closas[1,4], Jonas S. Almeida[1]

**Author affiliations**:
[1] Division of Cancer Epidemiology and Genetics, National Cancer Institute, Rockville, MD, USA.
[2] American Cancer Society, Atlanta, GA, USA.
[3] Bloomberg School of Public Health, Johns Hopkins University, Baltimore, MD, USA.
[4] The Cancer Epidemiology and Prevention Research Unit, The Institute of Cancer Research, London, England.

# Abstract

## Objective
Absolute risk models estimate an individual's future disease risk over a specified time interval. Applications utilizing server-side risk tooling, such as the R-based iCARE (R-iCARE), to build, validate, and apply absolute risk models, face serious limitations in portability and privacy due to their need for circulating user data in remote servers for operation. Our objective was to overcome these limitations.

## Materials and Methods
We refactored R-iCARE into a Python package (Py-iCARE) then compiled it to WebAssembly (Wasm-iCARE): a portable web module, which operates entirely within the privacy of the user's device.

## Results
We showcase the portability and privacy of Wasm-iCARE through two applications: for researchers to statistically validate risk models, and to deliver them to end-users. Both applications run entirely on the client-side, requiring no downloads or installations, and keeps user data on-device during risk calculation.

## Conclusions

Wasm-iCARE fosters accessible and privacy-preserving risk tools, accelerating their validation and delivery.

# INTRODUCTION

Absolute risk models estimate an individual's future disease risk over a specified time interval by utilizing information on known risk factors from healthy individuals in a population.[1] Individualized Coherent Absolute Risk Estimation (R-iCARE) is an R-based software package that facilitates building absolute risk models, validating them on independent studies that were not involved in model development, and applying them as a public health or clinical utility to estimate the absolute risk for an individual or a population.[2] R-iCARE has been used extensively to develop and validate models across both cancers and non-cancer outcomes.[3–21]

Applications developed using risk tooling written in programming languages that do not run natively on the web, such as those that depend on R, SAS, or Python,[2,22–28] are limited in terms of portability because they require maintaining a dedicated execution environment to run. To enable the application to run on various devices and operating systems, separate custom codebases need to be developed and maintained, often involving different native programming languages and technologies, while also ensuring that the risk calculation environment is compatible with each platform. It is challenging to set up the expensive infrastructure and expertise needed to support this.

The web is a ubiquitous and powerful platform to deliver software. The web is intrinsically portable as it enables applications developed using a single codebase to be distributed worldwide and be run through a web browser across various devices and operating systems, without the need for users to download or install anything. Web application codebases are developed using technologies in the core web stack: HTML, CSS, and JavaScript. The browser's internal secure sandboxed environment restricts web applications from accessing user data and system resources, thereby enforcing policies to protect users' privacy and security. The modern web reaches far beyond the confines of the browser, encompassing backend servers, serverless execution environments, the Internet of Things, and command-line tools;[29] with the advent of progressive web applications, it now provides an experience practically indistinguishable from device-native applications.[30] The reusability of modern web computing offers a particularly effective route to the FAIR principles (findable, accessible, interoperable, and reusable),[31] which guided the implementation reported here. Web applications that depend on non-web-stack codebases such as R-iCARE, for example using Shiny[32] or Plumber[33], require the development and maintenance of remote servers containing the necessary R-execution environment. For risk calculation, private user data is circulated on these servers, which are typically not under the user's governance. This is both a serious logistic and privacy concern: someone has to maintain and operate that server-side under stringent regulatory compliance mandates, and the user data circulates through opaque external environments.

In 2019, WebAssembly (Wasm)[34] became the fourth technology added to the core web stack. Wasm is a highly portable binary instruction format for a stack-based virtual machine that is designed to execute code at near-native speed in any environment with a Wasm runtime, both within and outside web browsers. Developers can now write parts of web application logic in programming languages other than JavaScript and then compile them into Wasm to enable it to be run directly on the web, without the need for external servers. This enables developers to leverage the existing libraries and communities of that language. At the time of writing, R's Wasm runtime environment is under active development but at a nascent stage.[35] JavaScript's data science ecosystem is rapidly growing but currently has limited libraries and community to support the requirements for risk calculation. Python's rich data science ecosystem and thriving community offers an attractive alternative. Pyodide, a Wasm-based Python distribution for the browser, can be used to run Python packages on the web.[36] iCARE re-written in Python and then compiled into Wasm can run directly on the web, inheriting all its benefits, including portability, without the need for external servers to support a Python-execution environment, and thereby preserving privacy since no data is circulated on opaque servers.

In this report, we describe the refactoring of R-iCARE into Python (Py-iCARE), and then consequently compiling it as a Wasm module (Wasm-iCARE) to enable iCARE to be run directly on the web. We showcase Wasm-iCARE, emphasizing its portability and privacy, through two applications: 1) for researchers to statistically validate risk tools, and 2) to deliver them to end-users. Both applications run entirely on the user's device, requiring no downloads or installations, and ensuring that no user data circulates outside their device.

# METHODS

The iCARE package can be used to build absolute risk models assuming that the age-specific disease incidence rates, given the risk factors, follow the proportional hazard model [37]. For a disease-free individual of age $a$, given their risk factor profile, $Z = \{z_1, \cdots, z_K\}$, the conditional incidence rate, $\lambda(a|Z)$, estimates their risk of developing the disease at age $a$. It is calculated using Equation 1. Here, the baseline hazard, $\lambda_0(a)$, gives the age-specific incidence rate when all risk factors take the reference values, and $\beta = \{\beta_1, \cdots, \beta_K\}$ is the log relative risk corresponding to each risk factor.

$$\lambda(a|Z) = \lambda_0(a) \cdot \exp(\beta^T Z) \qquad (1)$$

The absolute risk, $R$, is the probability that a disease-free individual of age $a$ will develop the disease over the time interval $[a, a + \tau]$. Using the model from Equation 1, we can calculate absolute risk using Equation 2, where $m(a)$ is the age-specific rate of competing events (e.g., mortality due to causes other than the disease of interest).

$$R_{a,\,a+\tau} = \int_{a}^{a+\tau} \lambda(t|Z) \cdot \exp\left(-\int_{a}^{t}\{\lambda(u|Z) + m(u)\} \cdot du\right) \cdot dt \qquad (2)$$

iCARE synthesizes standardized user-input data for model specification, including: 1) log relative risks (β), 2) marginal age-specific disease incidence rates, 3) a reference dataset representing the risk factor distribution of the underlying population, and 4) age-specific mortality rates from competing events. iCARE provides utilities for estimating baseline hazard from marginal disease incidence rates, imputing missing risk factors based on the reference dataset, and methods for model validation on an independent prospective cohort or a nested case-control study.[2]

To enhance iCARE's portability, we ensured the interoperability of the user-input data. Unlike R-iCARE, which operates exclusively in the R environment and uses the R Data Format (RDA), Py-iCARE works with common data exchange formats, such as, text, comma-separated values (CSV), and JavaScript Object Notation (JSON). This allows it to work across different programming environments, including R. To define the model, note the form of parametric nonlinear basis functions[38]: $f(\beta^\top \phi(X))$. It describes a set of parameterized, scalar-valued, nonlinear transformations called basis functions, $\Phi = \{\phi_1, \cdots, \phi_K\}$, of predictors, $X = \{x_1, \cdots, x_N\}$. They are linear with respect to the parameters (β) but can potentially be nonlinear with respect to the predictors ($X$). A function $f(\cdot)$ then performs some transformation of the weighted linear combination of the basis functions e.g., sigmoid in case of classification models. Each of linear regression, polynomial regression, and even neural networks can be seen as special cases of the basis function regression.[38] Users can specify either linear or nonlinear models of risk to plug into the exponent term of Equation 1. R-iCARE allows users to define these models using R's formula object. Py-iCARE uses a similar syntax with the Patsy package[39] to specify any basis function model through just a text input.

We then made the code itself portable by compiling Py-iCARE into WebAssembly (Wasm-iCARE) via the Pyodide package[36], to enable it to be run directly on the web.

# RESULTS AND DISCUSSION

## Model validation

Before risk models can be delivered to the end-user, researchers need to validate them on prospective cohorts. To demonstrate the use of Wasm-iCARE in model validation, we developed a validation module in a reactive web notebook, a reusable environment for real-time data manipulation, analysis, and visualization through dynamic code in a web-based interface.[40] The module validates the iCARE-Lit model on a simulated cohort study. iCARE-Lit[11,12] is a literature-based model to estimate the 5-year absolute risk of breast cancer. The model was formulated by synthesizing the relative risks reported in literature for self-reported risk factors, including: menstrual, reproductive, hormonal, family history, and lifestyle risk factors.

Figure 1 illustrates the architecture of this web notebook. When the researcher opens the notebook in a browser, the Wasm-iCARE module, iCARE-Lit model dependencies, and the validation dataset are loaded. Wasm-iCARE then validates the iCARE-Lit model against the dataset, returning statistical metrics for validation including model calibration and discrimination. As the red arrows in Figure 1 indicate, privacy is ensured because the validation data and calculations never leave the researcher's browser. In our demonstration, both the model dependencies and the validation dataset are publicly hosted, and therefore, anyone with access to the notebook can see and reproduce the results. The results can only be seen and reproduced by those with access to the notebook, model dependencies, and the same validation dataset. If, as shown in Figure 1, the researcher has privileged access to a restricted-access validation dataset in their private-server storage account, those without access to this validation dataset cannot run the analysis or see the results despite having access to the notebook. However, they may replicate the analysis on a different validation dataset that they have access to. To replicate this analysis on a different study, the researcher only needs to substitute the location (a dereferenceable URL) of the new validation dataset. Researchers execute the same code across validation studies, requiring no set up. This makes the code highly reusable without compromising on privacy. This demonstrates the conformity of Wasm-iCARE to the FAIR principles. The web notebook for model validation using Wasm-iCARE is provided at: https://observablehq.com/@jeyabbalas/icare-lit-model-validation.

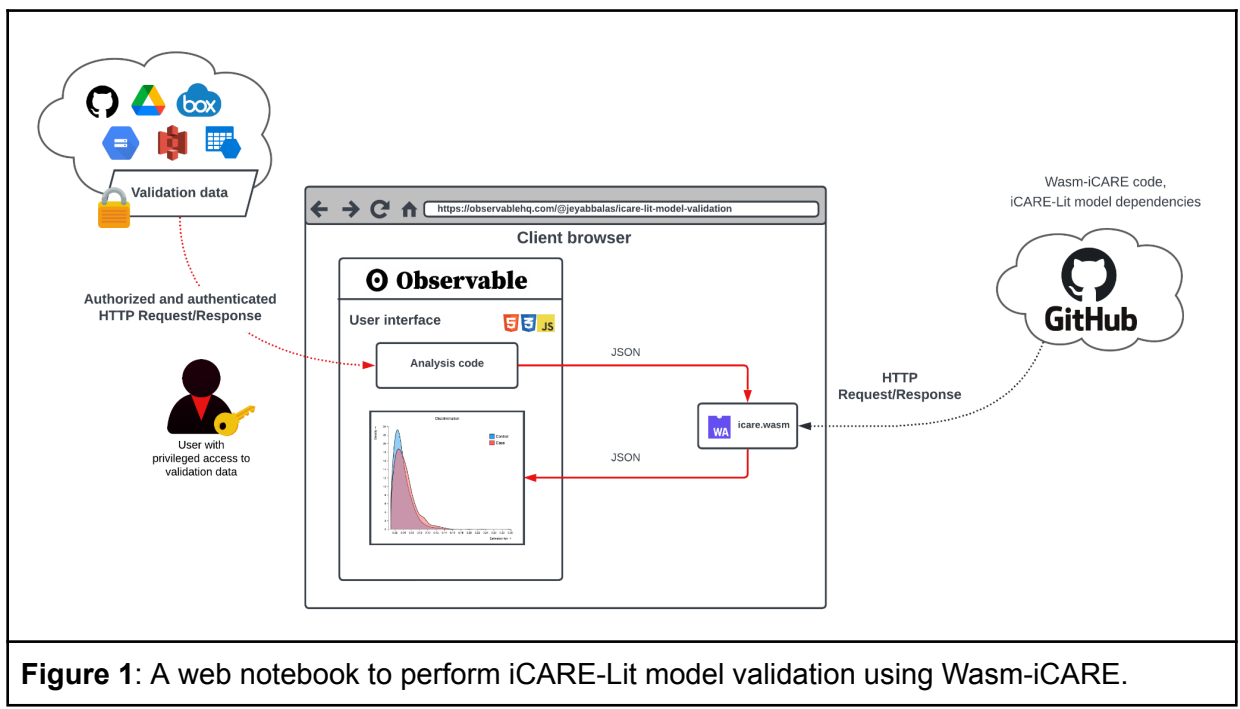

**Figure 1**: A web notebook to perform iCARE-Lit model validation using Wasm-iCARE.

## Model delivery

To demonstrate the use of Wasm-iCARE to deliver risk models, we used it to develop and distribute iCARE-Lit as a web application. Figure 2 illustrates its architecture. When a user opens the iCARE-Lit web application in a browser, the Wasm-iCARE module and the iCARE-Lit model dependencies are loaded. The user submits information on their risk factors by answering a questionnaire. The answers are submitted to the Wasm-iCARE module within their browser, which then calculates the 5-year absolute risk of breast cancer using Equation 2, and returns the results back to the user. Crucially, as the red arrows in Figure 2 indicate, the user data, the risk calculation, and the resulting risk assessment never leave the user's browser, ensuring privacy. This portable modular construct can be accessed and run on any device with a web browser, or any workflow engine operating in the web stack environment. Finally, users have the option to download their results locally for future reference, retaining full control over their data. The Wasm-iCARE powered iCARE-Lit application is publicly accessible at: https://jeyabbalas.github.io/icare-lit/.

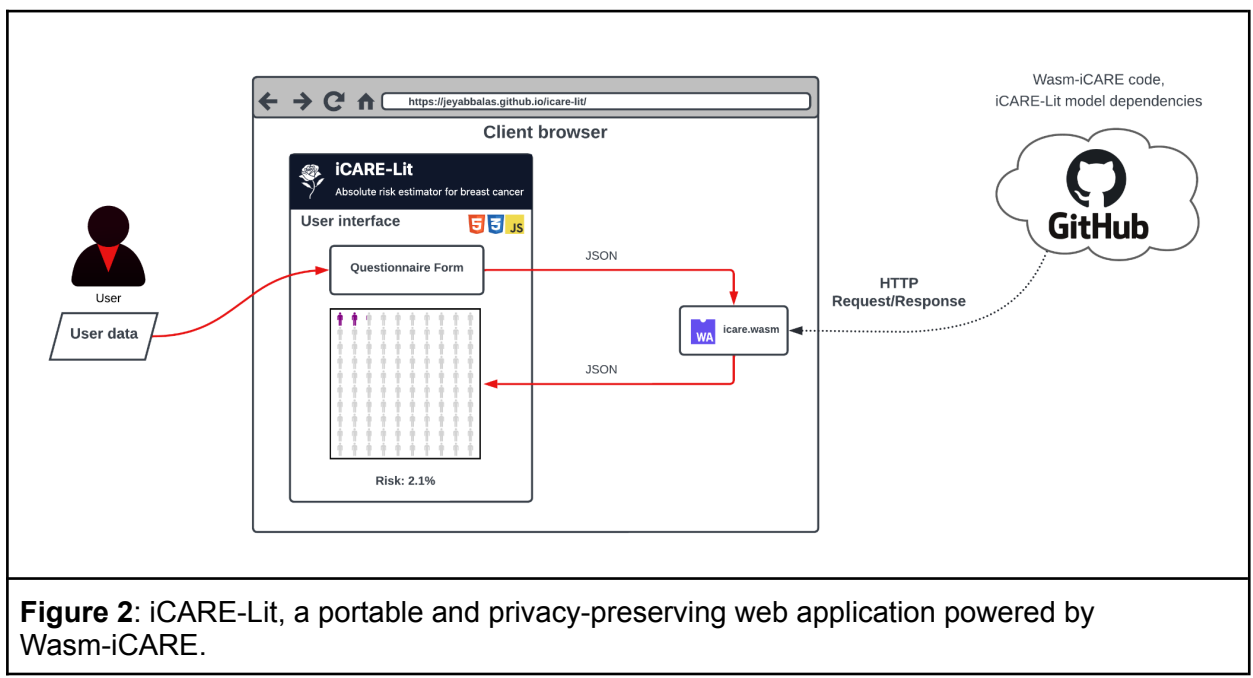

**Figure 2**: iCARE-Lit, a portable and privacy-preserving web application powered by Wasm-iCARE.

# CONCLUSIONS

We refactored R-iCARE into a Python package (Py-iCARE), and then compiled it into a WebAssembly (Wasm) module (Wasm-iCARE) to enable it to be run directly on the web. We demonstrated the portability and privacy of Wasm-iCARE for statistically validating risk models and delivering them to a user. Py-iCARE (https://github.com/jeyabbalas/py-icare) and Wasm-iCARE (https://github.com/jeyabbalas/wasm-icare) are open-sourced under the MIT license.

Wasm-iCARE moves risk assessment tooling to a more FAIR delivery model, which confers several benefits. Primarily, Wasm-iCARE runs risk calculations entirely on the user's device and under their governance, eliminating the need for web server hosting and maintenance. Consequently, there are no server costs, no network latencies during risk calculations, and applications using Wasm-iCARE can instantly scale to consumer demand. By allowing applications to move their code to the user's device instead of needing users to move their data, Wasm-iCARE can help comply with data privacy laws such as the General Data Protection Regulation (GDPR).[41] In a distributed computational setting, Wasm-iCARE facilitates model building via federated learning,[42] portable across client devices without their need to maintain a dedicated execution environment for risk calculation, and allows the communication of model summary statistics between data silos through web-stack real-time communication technologies such as WebRTC.[43]

Wasm-iCARE offers other server-dependent risk tools a software engineering blueprint to make them more accessible and privacy-preserving, to accelerate their widespread validation and distribution for advances in precision prevention.

# SUPPLEMENTARY MATERIAL

Developers who wish to utilize the iCARE package to build portable and privacy-preserving applications can use Wasm-iCARE to develop their applications. Wasm-iCARE is available at: https://github.com/jeyabbalas/wasm-icare. A series of web notebooks that demonstrates the use of Wasm-iCARE is available at: https://observablehq.com/@jeyabbalas/wasm-icare. For a demonstration of how to build web applications with Wasm-iCARE, visit: https://jeyabbalas.github.io/icare-lit/. Its source code is available at: https://github.com/jeyabbalas/icare-lit. Developers who wish to contribute additional functionalities to the iCARE library can do so at the Py-iCARE repository: https://github.com/jeyabbalas/py-icare.

# FUNDING

This work was funded by the National Cancer Institute (NCI) Intramural Research Program. PPC was supported by the Intramural Research Program of the American Cancer Society. NC was supported by NCI grant U01CA249866.

# ACKNOWLEDGEMENTS

The authors have no acknowledgments with respect to this publication.

# CONFLICT OF INTEREST STATEMENT

The authors declare no competing interests with respect to this publication.